\documentstyle[aps,twocolumn,epsfig]{revtex}

\newcommand{\beq}{\begin{equation}}
\newcommand{\eeq}{\end{equation}}
\newcommand{\bea}{\begin{eqnarray}}
\newcommand{\eea}{\end{eqnarray}}
\newcommand{\ba}{\begin{array}}
\newcommand{\ea}{\end{array}}
\newcommand{\bc}{\begin{center}}
\newcommand{\ec}{\end{center}}
\newcommand{\lsimeq}{\stackrel{<}{\scriptstyle\sim}}

\newcommand{\ie}{{\it i.e.}}
\newcommand{\eg}{{\it e.g.}}
\newcommand{\etal}{{\it et al.}}
\newcommand{\bml}{\begin{mathletters}}
\newcommand{\eml}{\end{mathletters}}
\newcommand{\half}{\hbox{$1\over2$}}
\newcommand{\commentout}[1]{{}}
\newcommand{\k}{{\bf k}}
\newcommand{\r}{{\bf r}}
\newcommand{\adag}{a^\dagger}
\newcommand{\bdag}{b^\dagger}
\newcommand{\gdag}{g^\dagger}

\newcommand{\lanl}{LANL e-print }

\begin{document}
\draft

\wideabs{
\title{Superposition of macroscopic numbers of atoms and molecules}
\author{John Calsamiglia$^1$, Matt Mackie$^1$, and Kalle-Antti
Suominen$^{1,2}$}
\address{$^1$Helsinki Institute of Physics, PL 64, 
  FIN-00014 Helsingin yliopisto, Finland \\
$^2$Department of Applied Physics, University of Turku, FIN-20014 Turun
yliopisto, Finland}
\date{\today}
\maketitle

\begin{abstract}
We theoretically examine photoassociation of a non-ideal Bose-Einstein
condensate, focusing on evidence for a macroscopic superposition
of atoms and molecules. This problem raises an interest because, rather than
two states of a given object, an atom-molecule system is a
seemingly impossible macroscopic superposition of different objects.
Nevertheless, photoassociation enables coherent intraparticle conversion, and
we thereby propose a viable scheme for creating a
superposition of a macroscopic number of atoms with a macroscopic number of
molecules. 
\end{abstract}
}

In a now famous {\it gedanken} experiment~\cite{FIRSTCAT},
Schr\"odinger highlighted the absurdity of applying quantum theory to
macroscopic objects by coupling the fate of a cat to the decay of a
radioactive atom, thereby forcing the animal into a superposition of
alive and dead. Over the years, this
nefarious situation has come to mark the fundamental
difference of principle between microscopic quantum mechanics and
macroscopic realism: whereas the latter asserts that a system with two (or more)
macroscopically distinct states must be in one or the other of these states, the former
of course allows for superpositions of different states~\cite{REALMAC}.
The technology of superconducting quantum interference
devices (SQUIDs) opened the door to the first proposed
macroscopic superposition~\cite{SQUIDCAT_T}, and the
latest SQUID experiments have taken a first step towards demonstrating such
a state~\cite{SQUIDCAT_E}. Other candidates for
creating macroscopic superpositions include nanoscale
magnets~\cite{NANMAG}, trapped ions~\cite{IONS},
photons in a high-$Q$ cavity~\cite{HIGHQ}, second-harmonic
generated~\cite{CHI2CATa,CHI2CATb} and amplitude-dispersed
photons~\cite{CHI3CAT}, C$_{60}$
molecules~\cite{BUCKY}, and two-component Bose-Einstein
condensates~\cite{LIGHTCAT,GRNDCAT,DYNOCAT} (BECs).
While it is a matter of contention as to whether or not these states are correctly
termed a Schr\"odinger cat~\cite{LOL}, their unambiguous observation would
nevertheless invalidate the concept of macroscopic realism.

Even more provocative, we propose coherent photoassociation of a
BEC~\cite{THEM,US} as a means of creating a
seemingly impossible macroscopic superposition of two different objects: atoms
and molecules. Photoassociation occurs when two atoms absorb a photon, thereby forming a
bound molecule. Besides the strongly enhanced efficiency of molecule
formation~\cite{THEMb,USb}, which actually relies on an order-unity
phase-space density rather than coherence~\cite{USb},
recent theoretical investigations into photoassociation of a BEC have led
to a variety of phenomena involving macroscopic quantum coherence.
For example, there are predictions of
coherent molecular solitons~\cite{THEM}, rapid~\cite{US} and
Raman~\cite{BSTIRAP} adiabatic passage
from an atomic to a molecular condensate,
Josephson-like BEC-MBEC oscillations~\cite{THEM,US,DJH},
and squeezed atomic condensates formed from an initial
MBEC~\cite{SQUEEZE}.

It is, however, a somewhat subtle point that marks the difference
between macroscopic coherence and a macroscopic superposition.
Given a collection of $N$ two-level systems, the macroscopic coherence that
leads, \eg, to Josephson-like oscillations, where all
$N$ systems tunnel in tandem between the two states, requires only that each
system is in a coherent superposition of the two levels, whereas a macroscopic
superposition with all of the systems in one or the other of the two states
necessarily exhibits an $N$-body coherence~\cite{LOL}. It is exactly this
distinction that warrants the present investigation.

Beyond the usual macroscopic superposition of two states of a given object,
photoassociation actually leads to a more counterintuitive situation since,
like (say) protons and quarks, molecules and atoms are different objects.
To the best of our knowledge, the only contemporary systems with a similar
potential are a BEC tuned to a Feshbach resonance~\cite{FESH}, which would
also give a macroscopic atom-molecule superposition, and second-harmonic
generation~\cite{SHG} (SHG), which would create, for example, a macroscopic
superposition of red and blue photons. These three systems are of course
mathematically identical, and the subsequent intramode superposition lies in the
possibility for coherent atom-molecule (red-blue photon) conversion. So far, SHG
has only yielded intramode phase superpositions between coherent
states~\cite{CHI2CATb,NOTE}, but these results are somewhat ambiguous as they
do not differentiate between a coherent superposition and a statistical mixture. In
contrast, the present work, based on unitary evolution of a Fock state, clearly
demonstrates an intramode superposition of a macroscopic number of atoms with a
macroscopic number of molecules. With photoassociation
currently~\cite{WYNAR} on the verge~\cite{OPUSMAG} of coherence, this state
should be realizable in the next generation of experiments.

Nonetheless, one-color photoassociation generally involves an excited
molecular state, and the ensuing spontaneous decay will inevitably introduce
decoherence. Moreover, rogue
photodissociation~\cite{US,OPUSMAG},
\ie, irreversible bound-free transitions to non-condensate
modes~\cite{ROGUEa,ROGUEb}, will destroy coherence as well.
In this Letter, we therefore consider an atom-molecule superposition formed {\it
via} two-color free-bound-bound photoassociation of an interacting BEC, where
the primary photoassociated molecules are transferred with another laser field to a
stable molecular state.  Adiabatic elimination~\cite{SCULL} of the intermediate
molecular state enables an effective two mode model involving only atoms and
stable molecules~\cite{DJH,OPUSMAG}, a sufficiently large
intermediate detuning should eliminate the harmful effects of irreversible  losses of
primarily photoassociated molecules, rogue~\cite{ROGUEa} or
otherwise~\cite{OPUSMAG}, and full inclusion of particle collisions-- itself a
novel aspect-- sets the table for the appearance of the desired superposition.

The development herein is outlined as follows. First, we review the
reduction of the problem from three to two modes, followed by an estimate
of the fixed parameters for which the neglect of irreversible decay is reasonable.
We then focus on a dynamical approach to creating a macroscopic superposition,
which begins with a joint atom-molecule condensate formed by the
photoassociative equivalent of a phase-tailored, strong $\pi/2$ pulse.  Switching
from a laser-dominated to a collision-dominated system, the symmetric
BEC-MBEC is then shown to evolve into a superposition comprised of
macroscopic numbers of atoms and molecules, with a ``size'' determined by the
ratio of the free-bound and intraparticle interactions. Finally, we discuss a possible
experimental signature, as well as the effects of any additional decoherence.

\begin{figure}
\centering
\epsfig{file=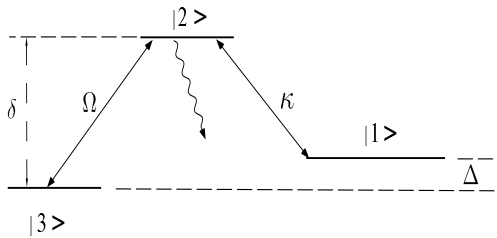,width=8cm,height=8cm}
\vspace{-1.3cm}
\caption{
Three-level illustration of coherent free-bound- bound
photoassociation, where $N$ atoms have assumedly Bose-condensed into state
$|1\rangle$. The free-bound and bound- bound  Rabi frequencies are $\kappa$ and
$\Omega$, respectively. Similarly, the two-photon and intermediate detunings
are $\Delta$ and $\delta$. The wavy line denotes the irreversible losses
that a large $|\delta|$ is expected to combat.
}
\label{THREEL}
\end{figure}

Turning to the situation of Fig.~\ref{THREEL}, we assume that $N$ atoms
have Bose-condensed into the same one-particle state, \eg, a plane wave
with wave vector $\k=0$. Photoassociation then removes two atoms from
this state $|1\rangle$, creating a molecule in the excited state
$|2\rangle$. Including a second laser, bound-bound transitions remove
excited molecules from state $|2\rangle$ and create stable molecules in
state $|3\rangle$. In second quantized notation, boson annihilation
operators for atoms, primarily photoassociated molecules, and stable
molecules are denoted, respectively, by $a$, $b$, and $g$.
The laser-matter interactions that drive the atom-molecule and
molecule-molecule transitions are written in terms of their respective
Rabi frequencies: $\kappa=d_1 E_1/2\hbar$ and $\Omega=d_2 E_2/2\hbar$,
where $E_i$ is the electric field amplitude, and $d_i$ is the corresponding dipole
matrix element $(i=1,2)$. Considering both
spontaneous and dissociative decay to occur to states outside those of
interest, these processes may be included with a non-Hermitian term
proportional to the total decay rate of the excited-molecular state~\cite{SCULL},
$\Gamma=\Gamma_s+\Gamma_d\,$. Defining the
two-photon and intermediate detunings as $\Delta$ and
$\delta$, the three-mode Hamiltonian is
then
\bea
H_3&=&-\Delta \gdag g + (\delta-\half i\Gamma)\bdag b
\nonumber\\
  &&-\half(\kappa\bdag aa + \kappa^*\adag\adag b)
    -(\Omega\gdag b + \Omega^*\bdag g),
\label{H_FULL}
\eea
where $\hbar=1$ and photon recoil effects are implicitly
included~\cite{US,OPUSMAG}.

First, we
consider the time evolution of the primary photoassociated molecules,
which is determined by the Heisenberg equation of motion
\beq
i\dot{b}=(\delta-\half i\Gamma) b -(\half\kappa aa+\Omega g).
\eeq
Assuming that $|\delta|$ is the largest evolution frequency
in the system, we set $\dot{b}/\delta=0$ and obtain
\beq
b\simeq\left({\half\kappa aa+\Omega g\over\delta}\right)
  \left(1+i\,{\Gamma\over2\delta}\right).
\label{B_APPROX}
\eeq
Substituting this result into Eq.~(\ref{H_FULL}) amounts to adiabatic elimination
of the excited molecular mode, and yields the effective two-mode
Hamiltonian~\cite{DJH,OPUSMAG}
\bea
H_2&=& -\Delta'\gdag g
    -\half(\chi\adag\adag g+\chi^*\gdag aa)
\nonumber\\
  &&+\lambda_1 \adag\adag aa + \lambda_2 \gdag\gdag gg
    +2\lambda_{12}\adag\gdag ag,
\label{H_EFF}
\eea
where the two-photon Rabi frequency is
$\chi=\kappa^*\Omega/\delta$ and the Stark-shifted two-photon detuning is
$\Delta'=\Delta +(|\Omega|^2/\delta)
  (1+i\,\Gamma/2\delta)$.
Additionally, we have introduced atom-atom, molecule-molecule,
and atom-molecule contact interactions with respective
coupling strengths
\bml
\bea
\lambda_1&=&{2\pi a_{11}\over mV_{11}}
   \left[1-{mV_{11}|\kappa|^2\over4\pi\delta a_{11}}
    \left(1-i\,{2\Gamma\over\delta}\right)\right],
\label{SHSL}
\\
\lambda_2&=&{2\pi a_{22}\over MV_{22}},
\\
\lambda_{12}&=&\lambda_{21}={2\pi a_{21}\over \mu V_{21}}\,,
\eea
\eml
where $a_{ij}$ is the particle-particle scattering length, 
$V^{-1}_{ij}=\int d^3r\,|\psi_i(\r)|^2|\psi_j(\r)|^2$ 
is the inverse effective volume of the single particle
wavefunctions ($i,j=1,2$),
$m$ ($M=2m$) is the mass of the atom (molecule),
and $\mu=2m/3$ is the reduced mass of the atom-molecule pair. 
Note that, due to the photoassociating 
laser~\cite{SL_SHIFT,DJH,OPUSMAG}, the atom-atom
scattering length in Eq.~(\ref{SHSL}) has also acquired a (complex) light shift.

Based generally on ratios of parameters, a quantum optics formalism is a definite
plus here since
the molecule-molecule and atom-molecule scattering lengths
are unknown. Hence,  we specify the magnitude of the intermediate detuning
as $|\delta|\gg\Gamma$, allowing for neglect of the
imaginary terms in $\Delta'$ and $\lambda_1$. To reduce the number of
parameters we also set $\lambda_2=4\lambda_1$, where the factor of four arises
from the conserved particle number being equal to the number of atoms plus twice
the number of molecules (see below). Exploiting the light-shifted scattering length, 
this equality could in practice be achieved by setting
${\rm sgn}\,\delta={\rm sgn}\,(a_{11}V^{-1}_{11}-2a_{22}V^{-1}_{22})$ and
$I_1/I_0=35.6947\,
  (V|a_{11}V^{-1}_{11}-2a_{22}V^{-1}_{22}|/\lambdabar)
    |\delta|/\epsilon_R\,$,
where $V$ is the BEC volume, $2\pi\lambdabar$ is the wavelength of the
photoassociating light,
$\epsilon_R=(2m\lambdabar^2)^{-1}$ is the photon recoil 
frequency, and 
we have used~\cite{OPUSMAG}
$\kappa/\epsilon_R=[(\lambdabar^3/V)(I_1/I_0)]^{1/2}$. The obscure numerical
factor comes from the definition of the
characteristic intensity $I_0$ for a given photoassociation 
transition~\cite{OPUSMAG}.
Focusing on $^{87}$Rb, we have~\cite{WYNAR,OPUSMAG}
$I_0=0.07\,$mW/cm$^2$,
$2\pi\lambdabar=780\,$nm, $\epsilon_R=3.77\times2\pi\,$kHz,
$\Gamma=12\times2\pi\,$MHz, and $a_{11}=5\,$nm.
Estimating 
$V|a_{11}V^{-1}_{11}-2a_{22}V^{-1}_{22}|=10\,a_{11}$, the fixed
parameters are $|\delta|\simeq10^3\times2\pi\,$MHz and
$I_1\simeq10\,$W/cm$^2$.

Within the two-mode approximation, the total number of particles $N=\adag a
+2\gdag g$ is conserved  and, without altering the physics, we may subtract the
term
$\lambda_1(N^2-N)$ from
$H_2$, resulting in the simplified form
\beq
H_2=H_0
    -\half\chi(e^{-i\varphi}\adag\adag g +e^{i\varphi}\gdag aa)
      -2\lambda\adag a \gdag g,
\label{H2M}
\eeq
where $H_0=-\Delta' \gdag g$, $\lambda=2\lambda_1-\lambda_{12}$, and a factor
of $\lambda_1$ has been absorbed into the detuning. Furthermore, $\chi$ is
hereafter a real quantity, and $\varphi$ specifies the relative phase between the
free-bound and bound-bound lasers. The main difference from the dual atomic
condensate problem~\cite{GRNDCAT,DYNOCAT} is that the
photoassociative coupling is trilinear,  and the ``tunneling rate", as opposed to
being $N$-independent, therefore scales as
$\sqrt{N}$.
Comparing $\sqrt{N}\chi$ to $N\lambda$,
the qualitative condition that determines a collision-dominated system is
$\chi/\lambda\lsimeq\sqrt{N}$, where the ratio $\chi/\lambda$ is readily adjustable
according to bound-bound intensity $I_2$.

\begin{figure}
\centering
\epsfig{file=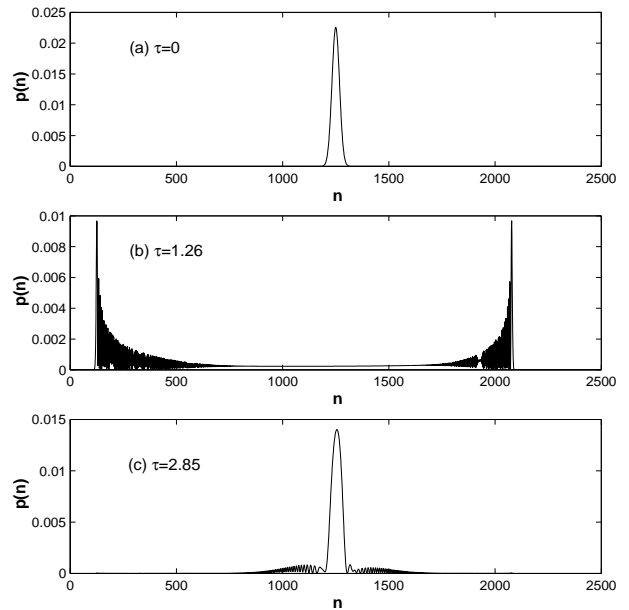,width=8cm,height=8cm}
\vspace{1cm}
\caption{
Creation of a macroscopic superposition of atoms and
molecules by photoassociating a BEC of atoms.
The frequency scale is defined by $\lambda=1$, the dimensionless time is
$\tau=\sqrt{N}\chi (t-\bar{t})$, and the initial state is  $|\psi_i\rangle=|N,0\rangle$
with $N=5000$ atoms and $n=0$ molecules.  (a) For $\varphi=\pi/2$, a strong and
two-photon resonant photoassociation pulse forms a symmetric joint
atom-molecule condensate with a real amplitude:
$|\psi(\tau=0)\rangle=\sum_{n=0}^{N/2} C_n(0)|N-2n,n\rangle$,
Im$[C_n(0)]=0$. Switching to
$[\varphi,\chi,\Delta']=[0,0.1\sqrt{N},0.1N\,]$,
the system evolves (b) into a coherent superposition of atoms and molecules
and (c) back to a symmetric atom-molecule condensate.
}
\label{FATCAT}
\end{figure}

Considering the formation of a macroscopic superposition of atoms and
molecules, we strongly suspect that a superposition as a ground
state~\cite{GRNDCAT} is absent due to the lack of particle-exchange
symmetry for the photoassociation Hamiltonian~(\ref{H2M}). Hence, we take a
dynamical approach~\cite{DYNOCAT}, which begins with the
production of a joint atom-molecule condensate having an ``evenly"
split population (\ie, $N/2$ atoms and
$N/4$ molecules), and a real-amplitude.
Setting the relative phase between the lasers to
$\varphi=\pi/2$, this state [shown in Fig.~\ref{FATCAT}~(a)] is formed by a
strong ($\chi/\lambda\gg\sqrt{N}$) two-photon-resonant photoassociation pulse
lasting a duration $\bar{t}$ determined by the solution to the semiclassical
approximation~\cite{SHG} for the number of molecules:
$N/4=(N/2)\tanh^2(\sqrt{2N}\chi\,\bar{t}\,)$. Interestingly enough, rather than
the binomial distribution of particles obtained for dual atomic
condensates~\cite{DYNOCAT}, the trilinear free-bound interaction
leads to a {\em squeezed} atom-molecule condensate. Such squeezing occurs
exactly as in SHG~\cite{CHI2CATa}, and is a source of many-particle
entanglement~\cite{SQUEEZE,MPENTANG}.

If, upon preparation of the dual atom-molecule condensate, the lasers are
adjusted so that~\cite{SYMCAT} $\varphi=0$,
$\Delta'\simeq\sqrt{N}\chi$, and $\chi/\lambda\lsimeq\sqrt{N}$, the
system will, as shown in Fig.~\ref{FATCAT}~(b), evolve
into a macroscopic atom-molecule superposition
over a timescale set by the Bose-enhanced two-photon Rabi coupling
$\sqrt{N}\chi$. The lack of smoothness in the probability distribution increases
with particle number, and occurs due to the nonlinear nature of the system.
Furthermore, the ``size" of the superposition, defined by the separation between the
main peaks, depends on the ratio
$\chi/\lambda$, and for a large number of particles is a maximum $\sim N/2$ when
$\chi/\lambda\simeq\half\sqrt{N}$. Consistent with the lack of
particle-exchange symmetry, the formation of this state depends on the relative sign
of $\chi$ and $\lambda$, which may be flipped if needed by setting $\varphi=\pi$.

Referring to Fig.~\ref{FATCAT}~(c), because the
system returns approximately to the initial dual
condensate, one could, for example, imprint a
phase on the molecular component of the system, leading to a signature
interference effect. Crucially, the rate of phase variations in the amplitude
$C_n(t)$, dictated by the collisional interaction term, is large only near
$n=N/4$ molecules, so that the subsequent decoherence from a temporally
imprecise imprinting of the phase should not effect a large superposition. The main
obstacle to realization is then external decoherence arising from interactions with
the ubiquitous thermal cloud of atoms; however, because the
BEC-MBEC superposition involves more than two states the loss of
coherence for one particle is not catastrophic. The
relevant frequency is $\sqrt{N}\chi=0.1N|\lambda|$, so that estimating
$\rho=10^{14}\,$cm$^{-3}$ and $|\lambda|=10\lambda_{\text{Rb}}$ leads
to a roughly millisecond birth rate. While the asymmetric
trilinear system will of course defy symmetrization, a trap engineered as a wide
well plus a narrow dip should sufficiently control the thermal-cloud
decoherence~\cite{NODECO}, leaving a superposition of macroscopic numbers of
atoms and molecules within reach of the next generation of photoassociation
experiments in $^{87}$Rb condensates.

We acknowledge Jani Martikainen and Jyrki Piilo for helpful discussions,
and Artur Ishkhanyan for pointing out the results of
Ref.~\cite{CHI2CATb}. This work was supported by
the Academy of Finland (project 43336) and the EU IST EQUIP program.

\vspace{-1.3cm}

\commentout{\newpage

\begin{figure}
\centering
\epsfig{file=THREEL.eps,width=8cm,height=10cm}
\end{figure}

\begin{figure}
\centering
\epsfig{file=FATCAT.eps,width=8cm,height=10cm}
\end{figure}
}

\begin{thebibliography}{99}

\bibitem{FIRSTCAT}
  E. Schr\"odinger, Naturwissenschaften {\bf 23}, 807; {\bf 23} 823; {\bf 23} 844
(1935).

\bibitem{REALMAC}
  A. J. Leggett, in {\it Quantum Measurement: Beyond Paradox},
    eds. R. A. Healey and G. Hellman (University of Minnesota Press,
      Minneapolis, 1998).

\bibitem{SQUIDCAT_T}
  A. J. Leggett and A. Garg, \prl {\bf 54}, 857 (1985).

\bibitem{SQUIDCAT_E}
  J. R. Friedman \etal,
    Nature {\bf 406}, 43 (2000).

\bibitem{NANMAG}
  J. R. Freidman \etal,
    \prl {\bf 76}, 3830 (1996);
  W. Wernsdorfer \etal, \prl {\bf 79}, 4014 (1997).

\bibitem{IONS}
  C. Monroe \etal,
    Science {\bf 272}, 1131 (1996).

\bibitem{HIGHQ}
  M. Brune \etal, \prl {\bf 77}, 4887 (1996).

\bibitem{CHI2CATa}
  S. P. Nikitin, and A. V. Masalov, Quant. Opt. {\bf 3}, 105 (1991).

\bibitem{CHI2CATb}
  S. T. Gevorkyan, \pra {\bf 62} 013813 (2000).

\bibitem{CHI3CAT}
  B. Yurke and D. Stoler, \prl {\bf 57}, 13 (1986).

\bibitem{BUCKY}
  M. Arndt \etal, Nature {\bf 401}, 680 (1999).

\bibitem{LIGHTCAT}
  J. Ruostekoski \etal, 
    \pra {\bf 57}, 511 (1998).

\bibitem{GRNDCAT}
  J. I. Cirac \etal, 
    \pra {\bf 57}, 1208 (1998).

\bibitem{DYNOCAT}
  D. Gordon and C. M. Savage, \pra {\bf 59}, 4623 (1999).

\bibitem{LOL} F. Lalo\"e, Am. J. Phys. {\bf 69}, 655 (2001).

\bibitem{THEM}
 P. D. Drummond, K. V. Kheruntsyan, and H. He, 
   \prl {\bf 81}, 3055 (1998). 

\bibitem{US}
  J. Javanainen and M. Mackie, \pra {\bf 59}, R3186 (1999).

\bibitem{THEMb}
  K. Burnett, P. S. Julienne, and K.-A. Suominen,
    \prl {\bf 77}, 1416 (1997);
  P. S. Julienne \etal,
    \pra {\bf 58}, R797 (1998).

\bibitem{USb}
  J. Javanainen and M. Mackie, \pra {\bf 58}, R789 (1998);
  M. Mackie and J. Javanainen, \pra {\bf 60}, 3174 (1999).

\bibitem{BSTIRAP}
  M. Mackie, R. Kowalski, and J. Javanainen,
    \prl {\bf 84}, 3803 (2000);
  M.~Mackie and J. Javanainen, \jmo {\bf 47}, 2645 (2000).

\bibitem{DJH}
  D. J. Heinzen \etal,  
    \prl {\bf 84}, 5029 (2000).

\bibitem{SQUEEZE}
  U. V. Poulsen and K. M\o lmer, \pra {\bf 63}, 023604 (2001).

\bibitem{FESH}
  P. Tommasini \etal,
    \lanl cond-mat/9804015;
  E. Timmermans \etal,
    \prl {\bf 83} 2691 (1999);
  F. A. van Abeelen and B. J. Verhaar, \prl {\bf 83} 1550 (1999);
  S. J. J. M. F. Kokkelmans, H. M. J. Vissers, and B. J. Verhaar,
    \pra {\bf 63}, 031601 (2001).

\bibitem{SHG}
  D. F. Walls and C. T. Tindle, J. Phys. A {\bf 5}, 534 (1972).

\bibitem{NOTE} The results of Ref.~\cite{CHI2CATa} pertain strictly to a
macroscopic intermode superposition.

\bibitem{WYNAR} 
  R. Wynar \etal, 
    Science {\bf 287}, 1016 (2000).

\bibitem{OPUSMAG}
  M. Ko\u{s}trun \etal, 
    \pra {\bf 62}, 063616 (2000).

\bibitem{ROGUEa}
  K. G\'oral, M. Gajda, and K. Rz\c{a}\.{z}ewski, 
    \prl {\bf 86}, 1397 (2001).

\bibitem{ROGUEb} M. Holland, J. Park, and R. Walser,
    \prl {\bf 86}, 1915 (2001).

\bibitem{SCULL} M. O. Scully and M. S. Zubairy, {\it Quantum Optics}
(Cambridge University Press, Cambridge, 1997).

\bibitem{SL_SHIFT}
  P. O. Fedichev \etal, 
    \prl {\bf 77}, 2913 (1996).

\bibitem{MPENTANG}
  A. S\o rensen \etal,
    Nature {\bf 409}, 63 (2001).

\bibitem{SYMCAT}
Due to the trilinear coupling, the phase diffusion process that leads to a cat tends
to favor the creation of atoms; hence, an initially symmetric
state ($N/2$ atoms, $N/4$ molecules) leads to an asymmetric superposition. A
detuning $\Delta'\simeq\sqrt{N}\chi$ helps balance the phase diffusion rate
between the particles, leading to roughly symmetric superpositions.


\bibitem{NODECO}
  D. A. R. Dalvit, J. Dziarmaga, and W. H. Zurek,
    \pra {\bf 62}, 013607 (2000).

\end{thebibliography}
\end{document}